\documentclass[aps,twocolumn,pre]{revtex4-2}

\usepackage[utf8]{inputenc}
\usepackage{graphicx}
\usepackage{xcolor}
\usepackage{soul}

\usepackage[mathlines]{lineno}
\usepackage[normalem]{ulem}
\usepackage{lipsum}
 \usepackage{amssymb,amsthm,amsmath}

\begin{document}

\title{
Critical habitat size of organisms diffusing with stochastic resetting
}

\author{Luiz Menon Jr.$^{1}$, Pablo de Castro$^2$,}
%\email{luizmenonjr@gmail.com}
\author{Celia Anteneodo$^{1,3}$}
\address{$^{1}$Department of Physics, PUC-Rio, Rua Marqu\^es de S\~ao Vicente 225, 22451-900, Rio de Janeiro, RJ, Brazil}
\address{$^2$ICTP-SAIFR,  Instituto de Física Teórica da UNESP, Rua Dr.\ Bento Teobaldo Ferraz 271, 01140-070, 
 S\~ao Paulo, SP, Brazil}
\address{$^{3}$Institute of Science and Technology for Complex Systems, INCT-SC, Brazil}

\begin{abstract}

The persistence of populations depends on the minimum habitat area required for survival, known as the critical patch size. While most studies assume purely diffusive movement, additional movement components can significantly alter habitat requirements. Here, we investigate how critical patch sizes are affected by stochastic resetting, where each organism intermittently returns to a common fixed location, modeling behaviors such as homing, refuge-seeking, or movement toward essential resources. We analytically derive the total population growth over time and the critical patch size. Our results are validated by agent-based simulations, showing excellent agreement. Our findings demonstrate that stochastic resetting can either increase or decrease the critical patch size, depending on the reset rate, reset position, and external environmental hostility. These results highlight how intermittent relocation shapes ecological thresholds and may provide insights for ecological modeling and conservation planning, particularly in fragmented landscapes such as in deforested regions. 

\end{abstract}

 \maketitle

\section{Introduction}
\label{sec:introduction}

 The critical patch size problem involves determining the minimum area required for a population of moving organisms to survive in a habitat surrounded by a hostile or unfavorable environment~\cite{macarthur2001theory,tilman1994competition,hanski1999metapopulation,fahrig2003}. As such, it is a fundamental concept in spatial population dynamics, crucial to understanding species persistence in fragmented landscapes~\cite{franklin2002habitat} and to developing strategies to mitigate its impacts~\cite{pimm2019,dickman2021,cantrell1999diffusion}. 
The critical size arises from the competition between population growth in the viable region and dispersal in the hostile environment~\cite{cantrell2001}.  
In its simplest form, the mathematical formulation of the problem amounts to considering population growth and diffusive spread across a boundary into an absorbing region \cite{kierstead1953size}. Its solution corresponds to finding the maximal eigenvalue of the Laplacian operator with Dirichlet boundary conditions. From this perspective, this question is crucial not only in population dynamics but also in both theoretical and applied contexts across a wide range of disciplines.

The critical size problem has been examined under various conditions, including inhomogeneous environments~\cite{COLOMBO2018,DOSSANTOS2020,LIN2004}, environmental conditions stochasticity~\cite{MENDEZ2010,Berti2015}, 
periodic oscillations~\cite{Ballard2004,COLOMBO2016}, 
and the presence of movement bias~\cite{Dornelas2024}, 
among others~\cite{Neicu2000}. 
These studies provide
valuable information on how different environmental factors influence population persistence. 
In fact, in real-world ecosystems, individuals often experience disruptions, such as environmental fluctuations or other forms of stochasticity, which can significantly impact their survival prospects. 
An important form of sudden disruption is the sporadic return of an organism to a preferred location or state in the system. One example are scenarios with predator-prey encounters \cite{mercado2018lotka}, which can be modeled by stochastic resetting~\cite{Evans_2011_prl,Evans_2020} of the individual's position. 
This reset can also be associated with occasional recolonization or migration and with behaviors such as central-place foraging~\cite{Bell1990}. The reset location can also represent resources such as a watering hole or a salt-lick.
 
Understanding how such resetting processes influence critical habitat sizes can provide insights into population persistence under uncertain conditions. In particular, stochastic resetting of position can allow an individual to overcome temporary resource scarcity or adverse conditions, potentially reducing the amount of space required for population survival. The existence of such a preferred location to which individuals return aligns well with field observations which show that individuals often have a tendency to concentrate around certain habitat regions~\cite{van2016movement}. In this study, we investigate the impact of stochastic resetting of position, taking as control parameters both the rate and position to which the individuals reset, on the critical habitat size of populations diffusing in one dimension. 

\begin{figure}[b]
    \includegraphics[width=1.\linewidth]{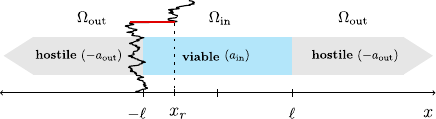}
    \caption{Pictorial representation of the one-dimensional (1D) spatial scenario used here. In the viable region, called patch or habitat, the growth rate is positive. This region extends from $x=-\ell$ to $x=\ell$. In the surrounding region, hostile conditions are present, implying a negative growth rate. The trajectory (time in the upward direction) depicts a random walk (black) with reset to position $x_r$ (red).}
    \label{fig:habitat}
\end{figure}

By examining the interplay between resetting dynamics and habitat properties, we aim to gain understanding about ecological resilience and population sustainability in more realistic scenarios. At the same time, by incorporating the stochastic reset mechanism, we aim to contribute to a broader framework in which the Laplacian problem is applicable, extending its relevance to more complex, real-world scenarios.

The model is defined in Section~\ref{sec:model}.
In Section~\ref{sec:extreme}, we examine the scenario under extreme external conditions, where the population decays at an infinite extinction rate $a_{\rm out}\to \infty$, which can be effectively modeled by setting absorbing boundary conditions. 
We also consider the general case with arbitrary $a_{\rm out}>0$, presented in Section~\ref{sec:arbitrary}.
In both scenarios, we explore how varying the reset rate and reset position influences the habitat size at which the population transitions to extinction, providing insight into the role of reset dynamics in population viability.
Final remarks and conclusions are presented in Section~\ref{sec:final}.

 \section{Model}
\label{sec:model}

In one dimension, we define a good-quality region---referred to as \textit{patch} or \textit{habitat}---as the interval $\Omega_{\rm in}=(-\ell,\ell)$. In the habitat, the population grows at a rate $a_{\rm in}$. This region is surrounded by a hostile environment (the complement set of $\Omega_{\rm in}$, denoted by $ {\Omega}_{\rm out}$), where the growth rate is negative,  $-a_{\rm out}$, as depicted in Fig.~\ref{fig:habitat}. 
Organisms move via normal diffusion plus sporadic random relocation from a pre-defined region $\Omega$ to a fixed position $x_r$ either inside or outside the habitat. For the region $\Omega$ from where reset occurs, two cases will be considered. The first will be the entire space $\Omega=\Omega_{\rm in}\cup\, \Omega_{\rm out}$ and the second just the outside region $\Omega=\Omega_{\rm out}$.
Moreover, we assume that the resets take place at random intervals, following a Poisson process with rate $r$, and each reset brings the walker to the fixed location $x_r$. 

Using a discrete, agent-based description, we have that the movement of an organism $i$ follows a Brownian motion supplemented with stochastic resetting, which can be mathematically described as follows:
\begin{eqnarray}
x^{(i)}(t+dt) &=&\left\{ \begin{array}{cl} 
    \stackrel{ 1- r dt}{=}  &  x^{(i)}(t) + \sqrt{2Ddt}\,\eta^{(i)}(t) ,   \\
    \stackrel{ r dt}{=} & x_r , 
\end{array} \right.  
\label{eq:mouvement_process}
\end{eqnarray}
where $D$ is the diffusivity, assumed to be the same everywhere, and $\eta^{(i)}$ is a Gaussian noise with zero mean and unit variance.
For population growth, we employ the reactions
\begin{align}
    i &\stackrel{a_{\rm in}\;\;}{\longrightarrow} i+i',  \;\;\;\; & \mbox{if $x\in \Omega_{\rm in},$\,\,\,} \nonumber  \\
     i&\stackrel{a_{\rm out}\;}{\longrightarrow} \varnothing, \;\;\;\;&  \mbox{if $x\in \Omega_{\rm out}$}, 
     \label{eq:growth_process}
\end{align}
where $i'$ is a new individual and $\varnothing$ denotes death.
Furthermore, we consider that at $t=0$ all walkers start at a position $x_0\in \Omega_{\rm in}$. 

Alternatively, the dynamics can be described in the continuum limit in terms of the density of organisms per unit length, $u(x,t;x_0)$, which represents the number of individuals per unit length at position $x$ and time $t$,
given an initial position $x_0$. Following the lines of Ref.~\cite{Evans_2011_prl,Evans_2011}, 
$u(x,t;x_0)$ is governed by the reaction-diffusion equation given by 

\begin{align}
\partial_t u &= \partial^2_x u + (1- \varepsilon \,r)u + r {N}_{\Omega} \delta(x-x_r),  & \mbox{if $x\in \Omega_{\rm in},\,\,\,$} \nonumber\\
\partial_t u &= \,\partial^2_x u - (a+r)u  + r{N}_{\Omega} \delta(x-x_r),                              & \mbox{if $x\in \Omega_{\rm out}$},
   \label{eq:u}   
\end{align} 
where $ u \equiv u(x,t;x_0)$, 
$N_{\Omega}(t,x_0) = \int_{\Omega} u\, dx$ and  
$\varepsilon=1$ (or 0) if $ \Omega$ contains $\Omega$ (or not). In addition, we rescaled the variables and parameters as follows:
 $t \,a_{\rm in} \to t$,
$x\sqrt{a_{\rm in}/D} \to x $,  
$a_{\rm out}/a_{\rm in} \to a$ 
($a_{\rm in}\to 1$), and
$r/a_{\rm in} \to r$. 
Note that in Eq.~\eqref{eq:u}, the term with the Dirac delta vanishes in one of the equations since $x_r$ belongs to one of the two regions only.  
 
In the equation for $x\in\Omega_{\rm in}$, 
the first term represents diffusion, the second term accounts for the growth at unit rate within the viable region together with the reduction due to organisms being relocated, while the last term gives the contribution from resets at position $x_r$,  which is nonzero only if $x_r \in \Omega_{\rm in}$.
In the equation for $x\in\Omega_{\rm out}$, 
the first term again represents diffusion, the second reflects the decay due to the negative rate $-a$ along with the removal of organisms via resetting, and the third term is the contribution from resets at $x_r$, which is nonzero only if $x_r \in \Omega_{\rm out}$.

Notice that we define the problem as linear from the outset, disregarding carrying capacity terms; however, our approach can be extended to more general scenarios (such as those involving logistic growth). Since we are primarily concerned with the critical condition near which the population density is very low, the carrying capacity will have no effect.

%\section{Results}
\section{Totally hostile environment}
\label{sec:extreme}

We first assume that, outside the good-quality patch, i.e., for $|x| > \ell$, one has that $a_{\rm out} \to \infty$, which means that organisms that leave the patch are instantly killed. Thus, only the first of the two equations in Eq.~(\ref{eq:u}) matters, i.e., the one for $x\in\Omega_{\rm in}$, with $\varepsilon=1$. The external region is mimicked by absorbing boundary conditions at $x=\pm \ell$.
Let us write the backward master equation associated to this process~\cite{gardiner1985handbook,Evans_2011}: 
 
\begin{equation}
    \partial_t u(x_0) = \partial^2_{x_0} u(x_0) + (1-r)u(x_0)  + r u(x_r),
    \label{eq:ubackward}
\end{equation}
where we denote $u(y)\equiv u(x,t;y)$.
Then, integrating over $x$, we obtain the following differential equation for the total number of organisms $N(t,x_0)$,  
\begin{equation}
    \partial_t N(t,x_0) = \partial^2_{x_0} N(t,x_0) + (1-r)N(t,x_0)  + r N(t,x_r).
    \label{eq:N_diff}
\end{equation}

 \begin{figure}[t!]
    \centering
\includegraphics[width=0.48\textwidth]{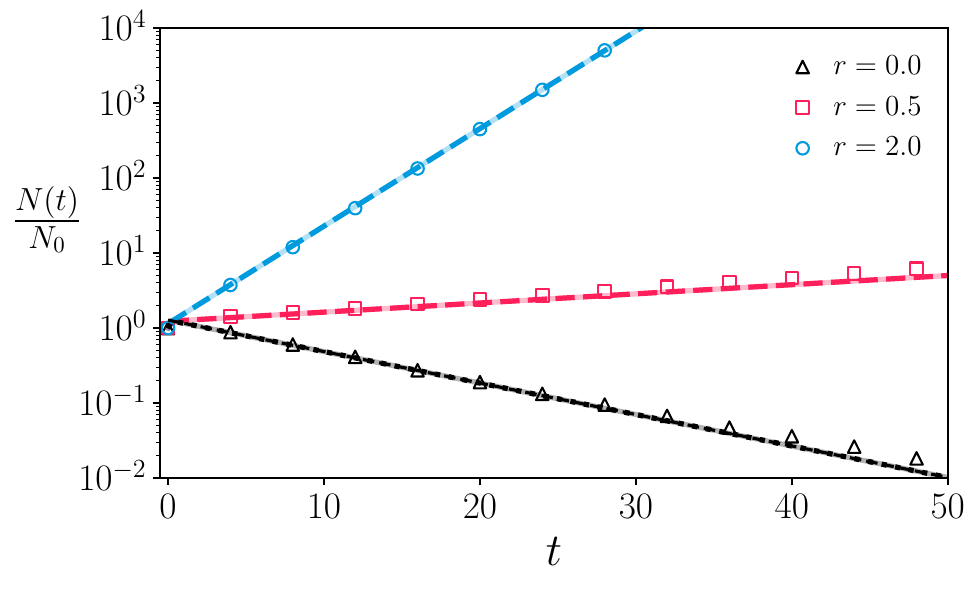}
    \caption{ Temporal evolution of the total population size $N(t)$ scaled by the initial values $N_0$, when the half width is $\ell=1.5$, $x_0=0$ and $x_r=0$ for given values of the resetting rate $r$ indicated in the legend. %
    The symbols refer to the average over $10^5$ stochastic simulations of Eqs.~(\ref{eq:mouvement_process}) and (\ref{eq:growth_process}) with $dt=10^{-5}$. 
    Thin solid lines are given by numerical inversion of Eq.~(\ref{eq:N_in_laplace})~\cite{hoog}, while dashed lines correspond to the long-time approximation provided by 
    Eq.~(\ref{eq:N}). 
 In the particular case $r=0$, we also plot (dotted line) the known closed-form expression~\cite{skellam1991random,kierstead1953size}, 
 and $s_M=s_{0} = 1-(\pi/3)^2$ (see Appendix~\ref{app:timesolution}).
    For $r>0$ the values of $s_M$ are obtained numerically and read $s_{0.5}\approx 0.0281$, $s_{2.0}\approx 0.2993$.
}
    \label{fig:evolution}
\end{figure}

The detailed steps for solving these equations are provided in the Appendixes. 
For sufficiently large times, we obtain
\begin{equation}
       N(t) \approx {\cal N}(x_0,x_r,r, \ell) \,e^{s_M t},
      \label{eq:N}
\end{equation}
where $\cal{N}$ is a constant that depends on all the parameters of the model, and $s_M$ is the dominant growth rate of a spectrum, which depends on $\ell$, $r$ and $x_r$ (see Appendix~\ref{app:timesolution} for details on the calculations).

In Fig.~\ref{fig:evolution}, we plot the evolution of $N(t)$ for different values of the reset rate $r$. 
Symbols correspond to numerical simulations of the stochastic process defined by Eqs.~(\ref{eq:mouvement_process}),  
while the dashed lines are given by the theoretical prediction provided 
by Eq.~(\ref{eq:N}),  in good agreement with the simulations.
In the case of Fig.~\ref{fig:evolution} ($\ell=1.5$ and $x_r=0$, which, in the absence of resetting, is slightly subcritical), we observe that the population decays, and the same happens for low reset rates. Conversely, for reset rates above a critical threshold ($r^*\simeq 0.4$), the population exhibits growth.

\subsection{Critical size}
\label{sec:criticalsize}

At the critical condition that separates the regimes of persistence and decline of the population, a stationary state corresponding to $s_M = 0$ emerges.
The critical size can be directly obtained by solving the stationary form of Eq.~(\ref{eq:N_diff}) (see  Appendix~\ref{app:steadysolution-infty}), leading to   
\begin{equation}
    \ell_c (r,x_r)= \frac{ \,\textrm{arccos}\left(r \cos \left[ x_r \sqrt{1-r}\right]\right)}{\sqrt{1-r}},
    \label{eq:criticalell-xr}
\end{equation}
where (and throughout the rest of the paper) the functions are defined for complex argument. 
The special (resonant) case $r=1$, for which the reset rate equals the growth rate, can be obtained by taking the limit $r\to 1$, which yields $\ell_c(1,x_r)=\sqrt{2+x_r^2}$.  
Setting $r=0$, we recover the well-known result in the absence of resetting: $\ell_c = \pi/2$.

\begin{figure}[b!]
    \centering
\includegraphics[width=0.45\textwidth]{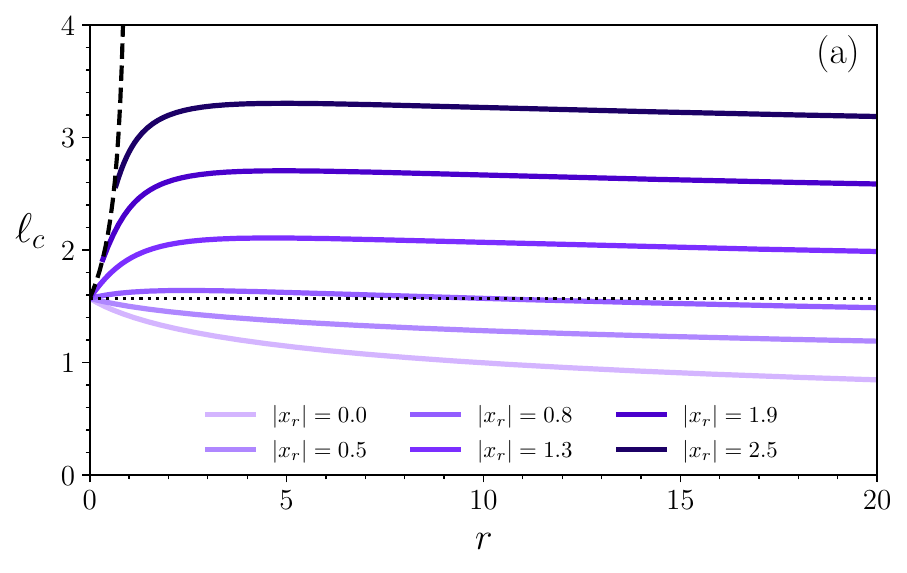}
\includegraphics[width=0.45\textwidth]{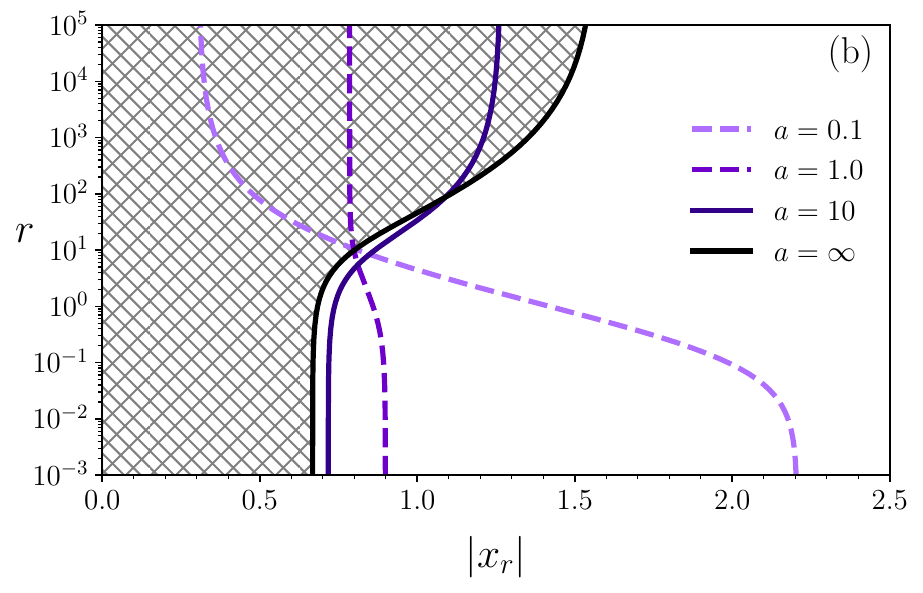}
    \caption{(a) Critical half-width $\ell_c$ as a function of resetting rate $r$ for different values of $|x_r|$ indicated in the legend. Solid lines are provided by Eq.~(\ref{eq:criticalell-xr}). 
    The horizontal dotted line corresponds to the critical value in the absence of resetting ($r=0$). 
    The dashed line is given by Eq.~(\ref{eq:out}). 
    (b) Diagram in the space of  parameters $(x_r,r)$ showing the (double-hashed) region where resetting reduces the critical patch size. 
    The black solid line corresponds to the finite $r^*$ for which
$\ell_c(r,x_r)=\ell_c(0,x_r)$ as a function of $x_r$. Beyond $|x_r|=\ell_c^*=\pi/2$, reduction of the critical size cannot be achieved for any value of $r$.
As explained in Section~\ref{sec:arbitrary}, colored lines represent $r^*$ vs.\ $x_r$ for finite values of $a$.}
    \label{fig:criticalline}
\end{figure}

Regarding the effect of the rate $r$ on the critical size for fixed reset position $x_r$, we present in Fig.~\ref{fig:criticalline}(a) a plot of $\ell_c$ vs.~$r$ for different values of $x_r$. Essentially, three different behaviors can be observed.

In particular, when $x_r=0$,  Eq.~(\ref{eq:criticalell-xr}) becomes
\begin{equation}
    \ell_c (r,0) = \frac{ \textrm{arccos}(r )}{\sqrt{1-r}},
    \label{eq:criticalell-0}
\end{equation}  
which decays with $r$ as shown in Fig.~\ref{fig:criticalline}(a). In the limit $r\to\infty$, one has $\ell_c\to 0$.
The same occurs for other reset positions $x_r$ provided that the slope $\partial \ell_c/\partial r|_{r=0} = \pi/4-\cos(x_r) <0$, i.e., for $|x_r| < x_r^*=\arccos(\pi/4) \simeq 0.667$. 
This range defines a core region around the center ($x=0$) where any resetting favors survival in a smaller habitat than in the reset-free case.

For $x_r$ sufficiently far from the center,  
stochastic reset cannot reduce the critical size regardless of the value of $r$.
In fact,  $\ell_c$ first increases with $r$ reaching a maximum and then decays to $\ell_c (\infty,x_r)$, which is equal to $|x_r|$, 
according to Eq.~(\ref{eq:criticalell-xr}). Therefore, 
for $ \ell_c(r=0)\equiv\ell_c^* < |x_r|$, habitat reduction cannot occur. 

In the intermediate region $x_r^*< |x_r| <  \ell_c^*$, 
although $\ell_c$ first increases with $r$, then, after reaching a maximum, it decays below the reset-free value $\ell_c^*$. This occurs provided that the reset rate is sufficiently high, namely, above the threshold rate $r^*$, at which $\ell_c(r^*,x_r)=\ell_c^*$. 
This corresponds to the point where each curve in Fig.~\ref{fig:criticalline}(a) intersects the dashed horizontal line. The threshold $r^*$ is plotted vs.~$|x_r|$ in Fig.~\ref{fig:criticalline}(b) (solid black line for $x_r\in[x_r^*,\ell_c^*]$). It indicates that, to prevent extinction, large rates are required when $x_r$ is located outside a core region around the habitat center. 
The double-hashed region is where the reset mechanism, characterized by $r$ and $x_r\in\Omega_{\rm in}$, allows a reduction in habitat size.

 Let us remark that a nonmonotonic behavior, as observed in Fig.~\ref{fig:criticalline} for $x_r$ sufficiently far from the boundaries, resembles that seen in scenarios where an elastic force guides organisms towards a specific area known as home range~\cite{Dornelas2024}. 
In that case, increasing the stiffness plays a similar role to increasing the reset rate in our problem. 
%which might be more closely emulated by a nonlocalized resetting. 
%
For weak forces, this effect can increase or decrease the critical patch size. However, when the force amplitude is sufficiently strong, organisms tend to concentrate in the return position.

Finally, let us remark that, when the reset position is outside the patch, $\ell_c (r,x_r) <|x_r|$, then, according to Eq.~(\ref{eq:u}), the effect of resetting is to effectively reduce the reproduction rate from $a_{\rm in}=1 \to 1-r$. This still allows survival for $r<1$ at a larger critical size given by the reset-free expression with $a_{\rm in}=1-r$:
\begin{equation} \label{eq:out}
    \bar{\ell}_c (r)=\frac{\pi}{2\sqrt{1-r}},
\end{equation} 
of course, independent of the reset position, since $x_r$ is in the absorbing region. 
This condition can occur only for $|x_r|>\bar{\ell}_c (0)=\pi/2$. In fact, setting $\bar\ell_c (r)=\ell_c (r,x_r)$, the non-null solution is $r=1-(\pi/[2x_r])^2$, plotted in Fig.~\ref{fig:criticalline}(a) as a dashed curve.

In summary: \\
(i) For $|x_r| < x_r^*$, the critical size is reduced by resetting at any rate $r$ (that is, $r^*=0$). \\
(ii) For $x_r^*<|x_r|< \ell_c^*$, the critical size can be reduced for rates $r>r^*>0$.  \\
(iii) For $\ell_c (0)<|x_r|$, the critical size cannot be reduced for any $r$.

\subsection{Stationary spatial distribution at criticality}

\begin{figure}[b!]
    \centering
\includegraphics[width=0.4\textwidth]{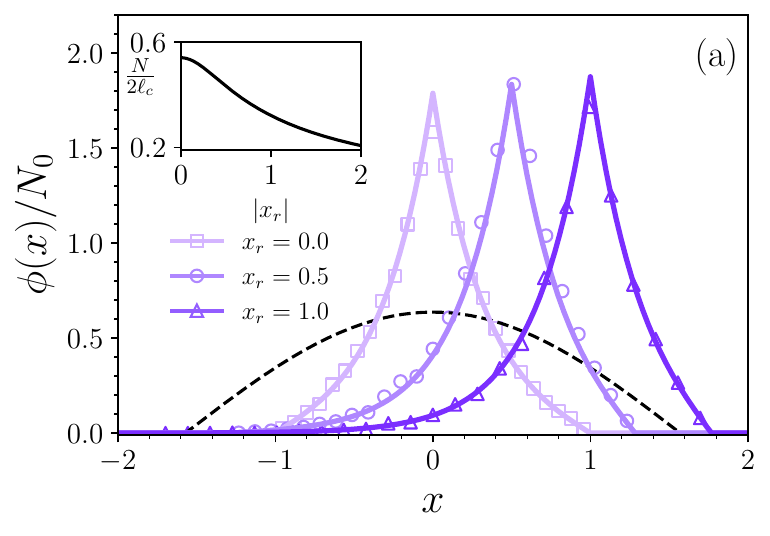}
\includegraphics[width=0.4\textwidth]{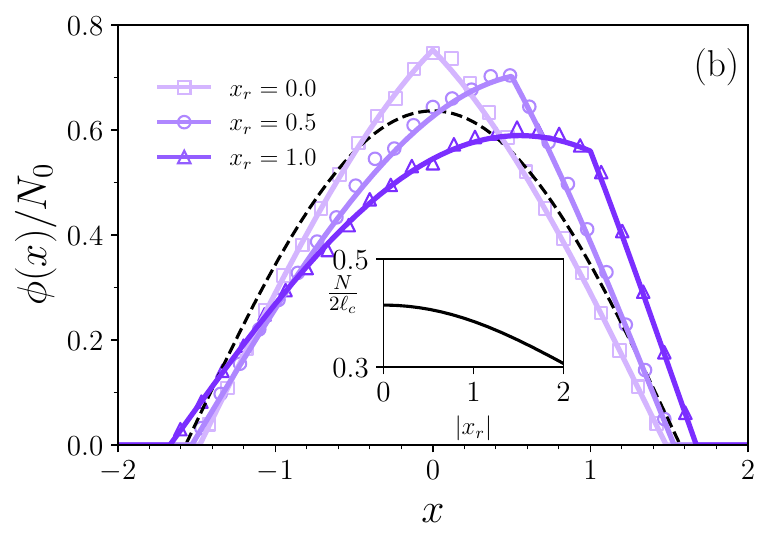}
    \caption{Stationary density distribution for $r=10$ (a) and $r=0.5$ (b), $x_r= 0.0$, 0.5 and 1.0, at the critical values of the patch size $2\ell_c$ (see Fig.~\ref{fig:criticalline}(a)). The dashed curve corresponds to the stationary state for $r=0$, drawn for comparison. In all cases,   $a_{\rm out}\to\infty$. The  lines are provided by Eq.~(\ref{eq:stationary}) and  the symbols are provided by stochastic simulation of Eq.~(\ref{eq:mouvement_process}) with $10^5 $ realizations and $dt= 10^{-5}$. The insets display the stationary value of the total population per unit length vs.\ $|x_r|$ when the initial density is $N_0=1$.
    }
    \label{fig:density}
\end{figure}

For a sufficiently long time, when the dominant mode has settled (which can occur very quickly, as observed in Fig.~\ref{fig:evolution}),  we can approximate $u(x,t) \approx \phi(x)\exp(s_M t)$, which substituted into Eq.~(\ref{eq:u}) gives 
\begin{equation} \label{eq:phi}
    \partial^2_x \phi + \left(1-r-s_M \right)\phi =- r {\cal N} \delta(x-x_r),
\end{equation}
where ${\cal N}\equiv {\cal N}(x_0,x_r,r, \ell)$ was defined in Eq.~(\ref{eq:pre_factor0}). 
The solution of Eq.~(\ref{eq:phi}) is the Green function associated to the 1D Helmholtz equation, that is, 
 \begin{equation}
 \label{eq:stationary}
    \phi(x)= r {\cal N}
\begin{cases} 
    \displaystyle{ \frac{   
     \sin\left[ \xi_r\,(\ell + x)\right]\sin\left[\xi_r\,(\ell - x_r)\right]}{\xi_r \sin[( 2\xi_r\,\ell]}}  , & -\ell < x < x_r,\\[5mm]
   \displaystyle{ \frac{ \sin\left[ \xi_r\,(\ell - x)\right ]\sin\left[  \xi_r\,(\ell + x_r)\right]}{\xi_r \sin[ 2\xi_r\,\ell ]}} , & x_r < x < \ell,
\end{cases}
\end{equation}
where $\xi_r = \sqrt{1-r-s_M}$ (see Appendix \ref{app:timesolution}). 
The initial assumption implies that the spatial profile tends to a shape that remains unchanged except for the exponential pre-factor.

At criticality, where $\ell=\ell_c$ and $s_M=0$, $\phi(x)$ coincides with the stationary solution. 
These theoretical
stationary profiles are shown in Fig.~\ref{fig:density} (lines), for $r=10$ (a) and $r=0.5$ (b), and different values of $x_r$, in good agreement with the agent-based simulations. 
Note again that, for reset positions near the center (small  $|x_r|$), the reset mechanism reduces the critical habitat size. However, this effect is not observed for positions closer to the borders. Furthermore, the stationary value of the total population density, $N/(2\ell_c)$, decreases with increasing $x_r$, as illustrated in the insets.

\begin{figure}[h!]
    \centering
\includegraphics[width=0.4\textwidth]{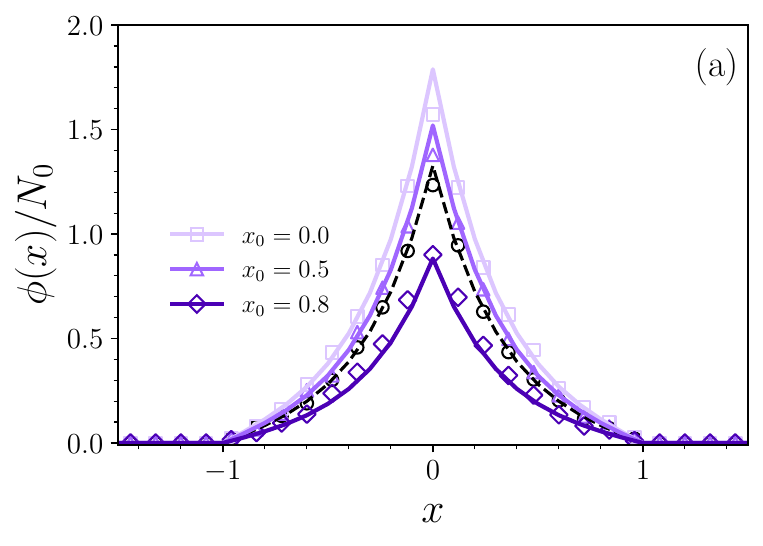}
\includegraphics[width=0.4\textwidth]{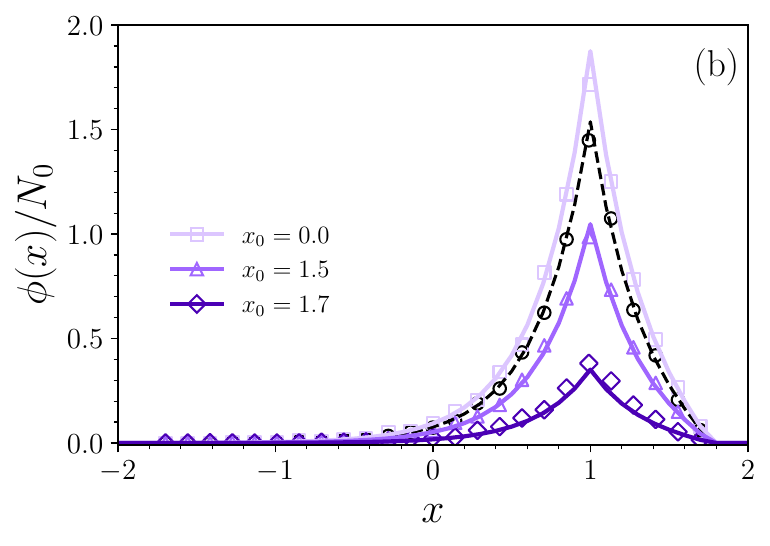}
    \caption{Stationary density distribution, when the initial distribution is a Dirac delta centered at $x_0$, 
   for different values of $x_0$ indicated in the legend. Also, $r=10$ and the reset positions are  (a) $x_r=0$  and  (b) $x_r=1$. 
    The case where the initial distribution is uniform is also plotted for comparison (black dashed line). 
     }
    \label{fig:ic}
\end{figure}

\subsection{Role of the initial distribution}
  \label{sec:initial}

  Recall that the initial condition considered is a Dirac delta function at an arbitrary point $x_0$ in the high quality region, but Eq.~(\ref{eq:criticalell-xr}) for the critical patch size remains independent of this initial concentration point.

 Fig.~\ref{fig:ic}(a) shows the steady states resulting from initial conditions concentrated at different initial points $x_0$, for the reset position $x_r=0$ and $r=10$. Note that the shape is invariant but its height decreases when the reset position is shifted towards the borders, as expected due to larger diffusive losses during the transient. The case of uniform initial distribution is also plotted for comparison, yielding also the same stationary profile except for a multiplicative factor. Also in this case the critical habitat size is unchanged. 
 In fact, since any general initial distribution can be represented as a continuous superposition of Dirac delta functions, and the evolution equation is linearized, then, in particular, the critical size is unaffected by the choice of initial distribution, as illustrated by the stationary spatial profile that emerges from the uniform distribution. Fig.~\ref{fig:ic}(b) illustrates the effect of the initial condition for an off-center resetting position ($x_r=1$), displaying qualitatively similar effects to those in case (a).

\section{Partially hostile environment}
\label{sec:arbitrary}

We now consider scenarios in which the region outside the patch of size $2\ell $ is not completely degraded, but instead has finite extinction rate $a\in (0,\infty)$, while stochastic returns can occur from a given region $\Omega$ to a position $x_r$ within or outside the patch. We analyze two main relevant cases, for which we find the critical habitat size. One of them is when $\Omega=\Re$, that is, organisms both inside and outside the habitat can be reset, what we call ``total relocation''. Another is when $\Omega= \Omega_{\rm out}$, that is, only the organisms in the external region can be reset, what we name ``partial relocation''.

%%%%%%%%
\subsection{Total relocation }
\label{subsec:all2*}

Organisms inside and outside the patch can be selected for sudden relocation. Then, Eqs.~(\ref{eq:u}) become
\begin{align}
 & u'' +   (1 -r)u + r \,N \delta(x-x_r)=0,  & \quad & \mbox{if } x\in \Omega_{\rm in},\nonumber\\
 &   u'' - (a+ r)u+r N \,\delta(x-x_r)=0,                              & \quad & \mbox{if } x\in \Omega_{\rm out}.
   \label{eq:all2*}   
\end{align}
If $x_r\in \Omega_{\rm in}$, that is, if organisms are relocated to the patch, we obtain (following the approach described in Appendix~\ref{app:steadysolution-all}), 
\begin{equation}
    \ell_c(r,x_r)=   \frac{1}{\sqrt{1-r}}\arccos
    \Biggl(
 r  \cos(\alpha x_r) + \sigma 
\sqrt{\Lambda} 
    \Biggr),
    \label{eq:criticalell-all2in}
\end{equation}
where 
$$\Lambda=
 (1-r)\Bigl[\frac{1}{a+1} -   \frac{r^2 \cos^2(\alpha x_r)}{a+r}\Bigr],
 $$
with $\alpha=\sqrt{1-r}$ and $\sigma={\rm sign}(1-r)$.

If $x_r \in \Omega_{\rm out}$, that is, if organisms are relocated to a point outside the patch, then a transcendental equation for the critical size is obtained, namely, 

\begin{align}
\sin (\alpha\,  \ell_c) \left[(a+1)\, r\, e^{\gamma  \ell_c}+\alpha ^2 \left(\gamma ^2-r\right) e^{\gamma  x_r}\right]&\nonumber\\
- \alpha  \gamma   \left(\gamma ^2-r\right) e^{\gamma  x_r} \cos
   (\alpha \, \ell_c)&=0,
    \label{eq:condition-all2out}
\end{align}
where $\gamma=\sqrt{a+r}$. 
%The numerical solution is used in Fig.~\ref{fig:finite-all}(b) (dashed line).

\begin{figure}[t]
	\centering
	\includegraphics[width=0.45\textwidth]{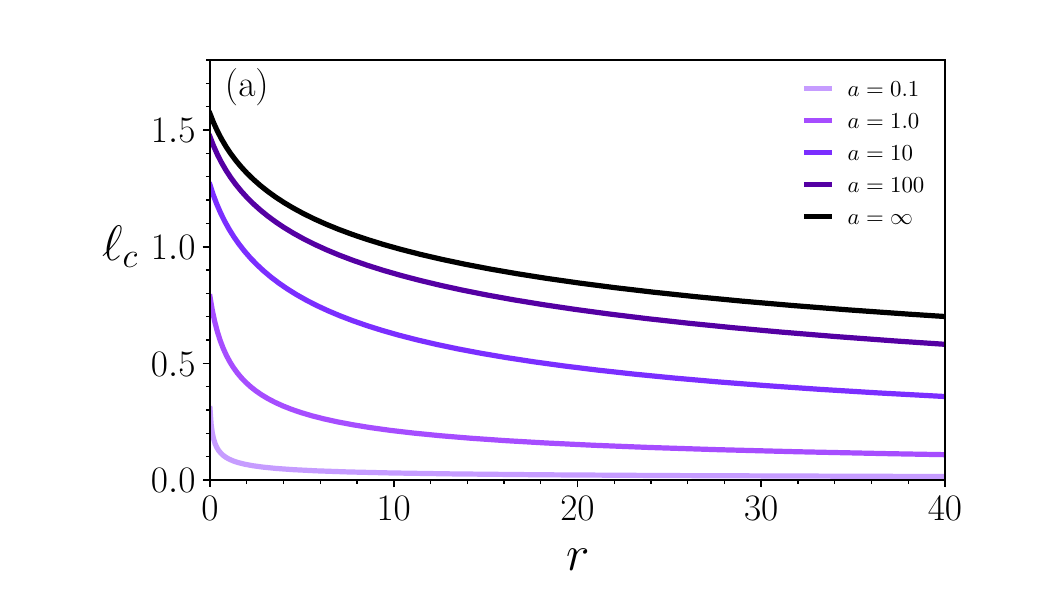}
	\includegraphics[width=0.45\textwidth]{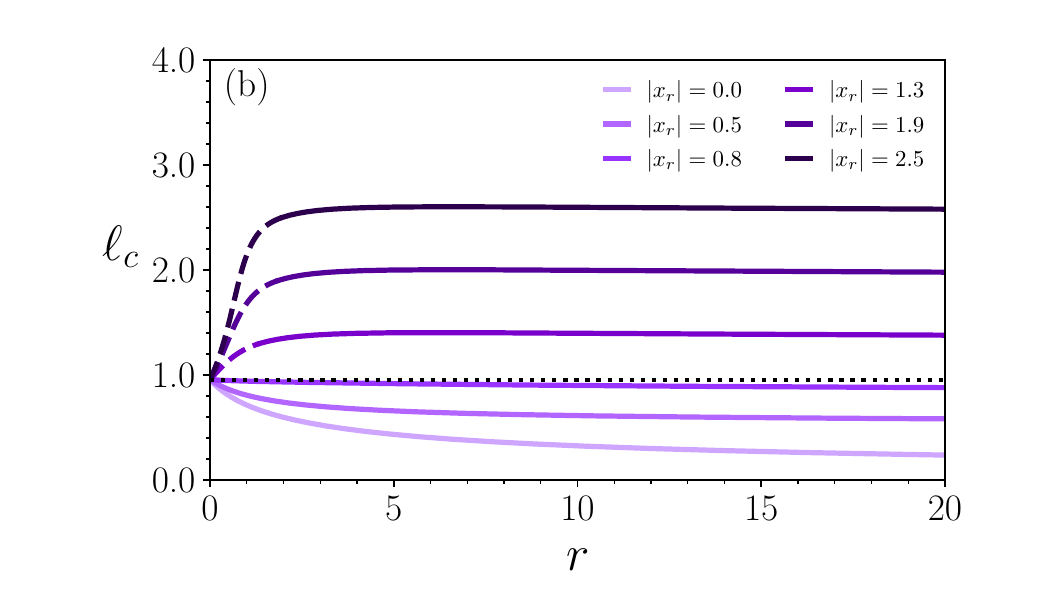}
	\includegraphics[width=0.45\textwidth]{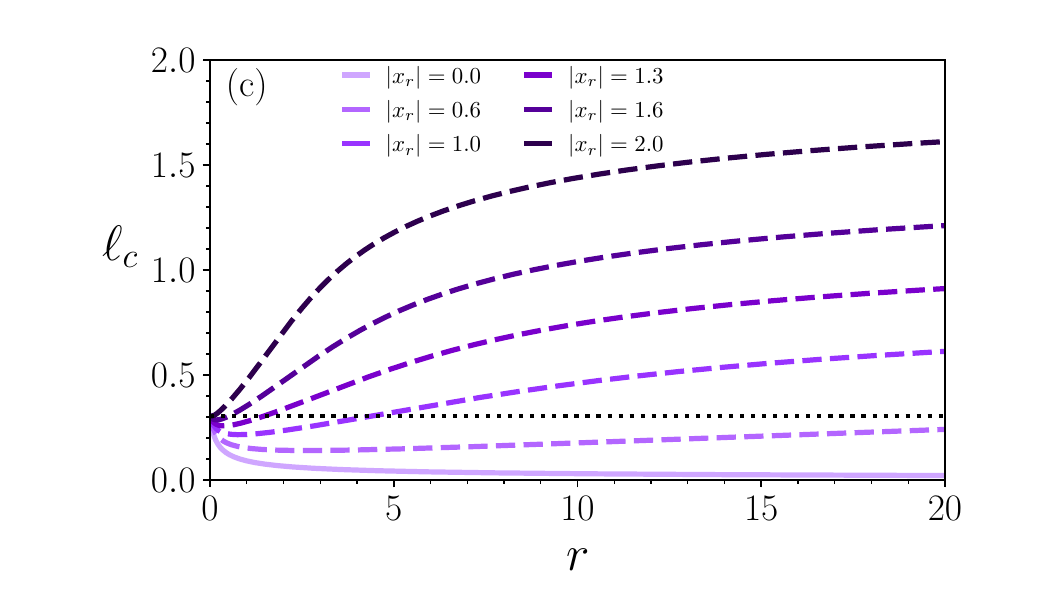}
	\caption{Repositioning from any location.     Critical half-width $\ell_c$ as a function of rate $r$, (a) for different values of $a$ when $x_r=0$, and for $a=2$ (b) and $a=0.1$ (c) for different values of $|x_r|$. 
		Solid lines are provided by Eq.~(\ref{eq:criticalell-all2in}), for $x_r\in \Omega_{\rm in}$. Dashed lines are obtained by numerically solving Eq.~(\ref{eq:condition-all2out}), with $x_r\in \Omega_{\rm out}$.
		In (b), the dotted horizontal line represents the critical half-width in the absence of resetting.
	}
	\label{fig:finite-all}
\end{figure}

Figure~\ref{fig:finite-all}(a) shows the critical half-width 
$\ell_c$ as a function of $r$, for the special case $x_r=0$ and different values of $a$. In the absence of resetting, less hostile environments (i.e., lower decay rate $a$) lead, of course, to a reduced critical size. 
When reset is introduced and $x_r=0$, then the increase in the reset rate $r$  monotonically reduces $\ell_c$, as expected since more frequently organisms are conducted to the most favorable point. This effect is more pronounced for smaller values of $a$. In particular, when $a$ approaches zero, the critical size tends to zero as well. In the opposite limit $a\to \infty$, we recover the curve for $|x_r|=0$ shown in Fig.~\ref{fig:criticalline}(a). 

For other values of $x_r$, the curves $\ell_c$ vs. $r$ can exhibit several different nonmonotonic behaviors, as shown in Figs.~\ref{fig:finite-all}(b)-(c). For $a=2$ (b), the picture is qualitatively similar to that observed for the totally hostile environment in Fig.~\ref{fig:criticalline}, with the three types of regime described at the end of Sect.~\ref{sec:criticalsize}. We remark that the dashed line for $|x_r|=1.6$ was obtained by numerically solving Eq.~(\ref{eq:condition-all2out}), which recapitulates Eq.~(\ref{eq:out}) when $a\to \infty$.
In this limit, the full portrait provided by Fig.~\ref{fig:criticalline}(a) is recovered.

%%%%%%%%%%%%%%%%%%%%%%%%%%%%%%%%%%%%%%%%%%%%%%%%%%%

\begin{figure}[t!]
    \centering
    \includegraphics[width=0.45\textwidth]{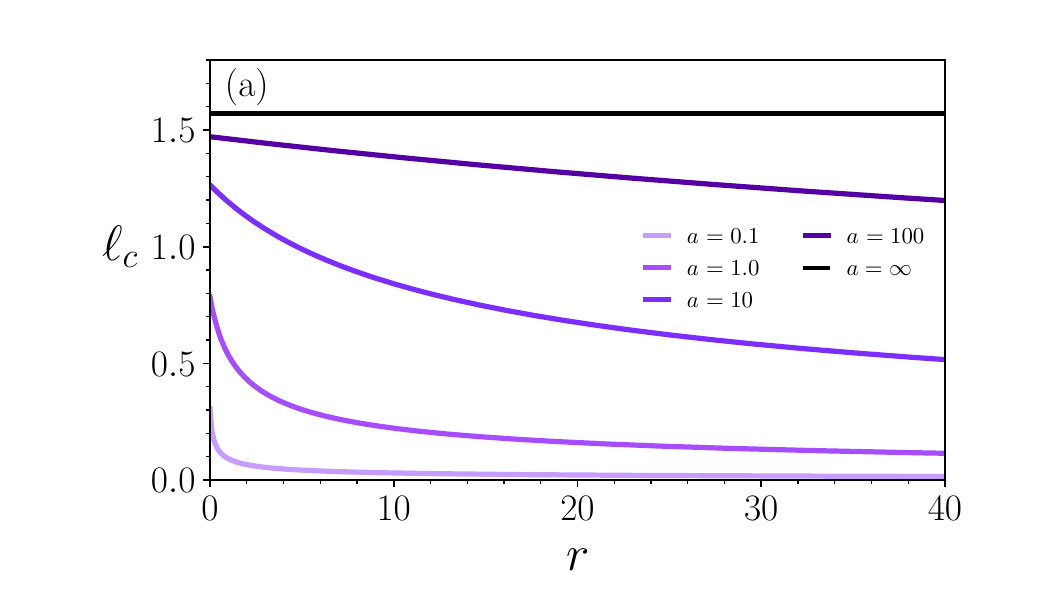}
    \includegraphics[width=0.45\textwidth]{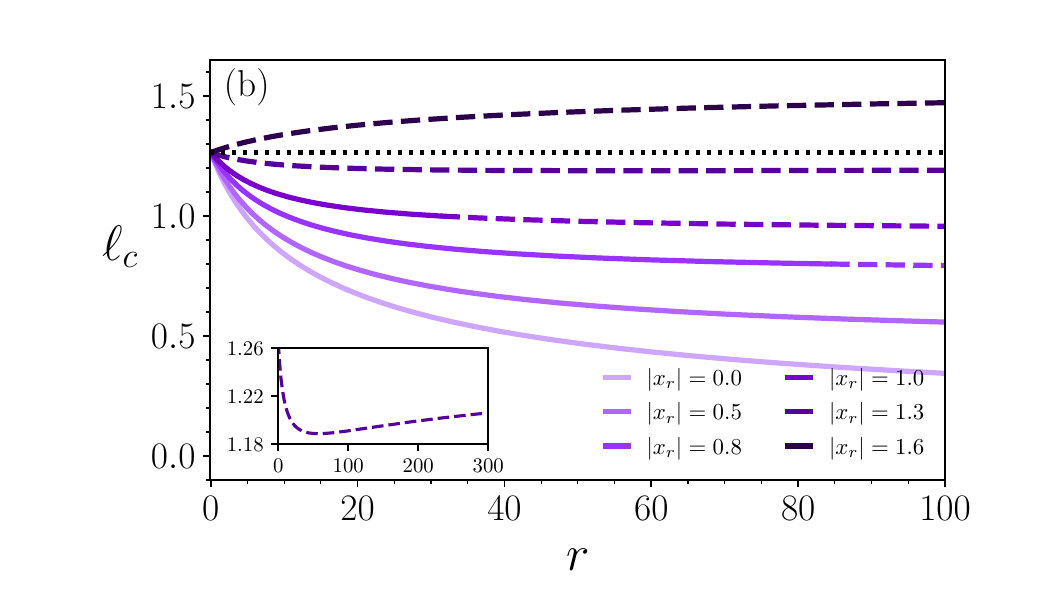}
    \includegraphics[width=0.45\textwidth]{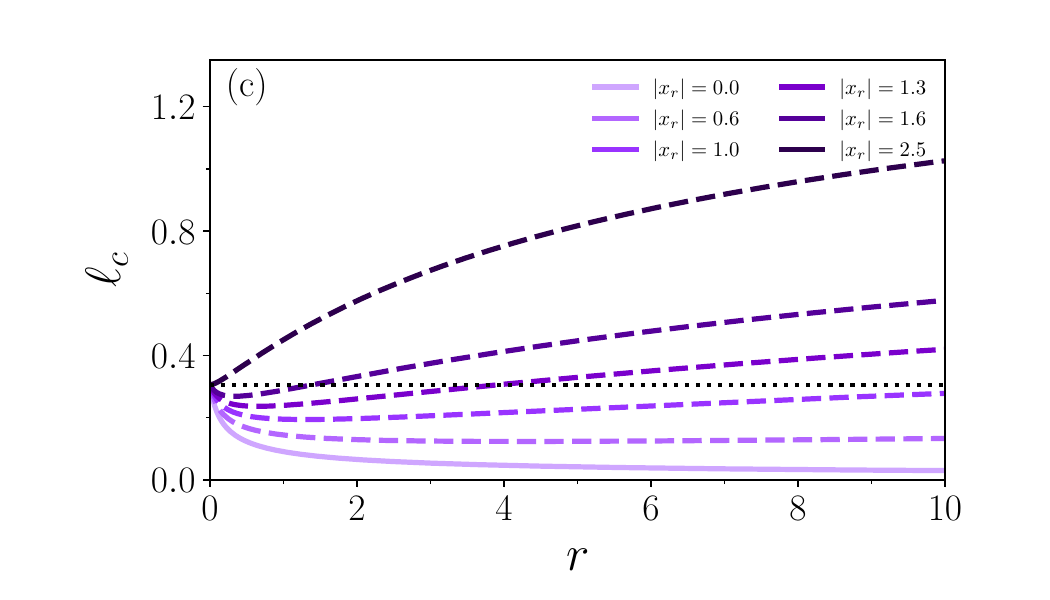}
    \caption{Repositioning from outside the patch only.     
    Critical half-width $\ell_c$ as a function of rate $r$, (a) for different values of $a$, with $x_r=0$, 
    for $a=10$ (b) and $a=0.1 $(c) with different values of $|x_r|$. 
Solid lines are provided by Eq.~(\ref{eq:criticalell-out2in}), for $x_r\in \Omega_{\rm in}$. Dashed lines are obtained by numerically solving Eq.~(\ref{eq:condition-out2out}),  for $x_r\in \Omega_{\rm out}$. 
 In (b), the dotted horizontal line represents the critical half-width in the absence of resetting.
 The inset shows a close-up of the case $x_r=1.3$.
 }
    \label{fig:finite-out}
\end{figure}

However, when the environment is weakly hostile (i.e., $a$ is small), the picture departs from that of the totally hostile environment, as illustrated in 
Fig.~\ref{fig:finite-all}(c), for $a=0.1$. 
Typically, for small enough $x_r$, there is an optimal (finite) reset rate, which produces a minimal critical size (hence smaller than the reset-free one)  
for resetting positions outside the habitat (dashed lines),
as illustrated for $a=0.1$ in Fig.~\ref{fig:finite-all}(c).
This effect can be qualitatively understood as follows. 
For a highly hostile environment (large $a$), resetting becomes detrimental when the reset position is located far from the patch center, even if it is still within the patch. In such cases, survival requires an increase in habitat size.
Differently, in a weakly hostile environment  (small $a$), there can be a positive balance of individuals that benefit from being reset to a position close to the patch, even if outside the patch, so that they do not spend too much time well within the outside region.  In such case, the reset dynamics can lead to a reduction in the minimal habitat size. 
This balance involves the timescales of the reset, growth and diffusive processes. 
However, if the reset position is too distant or the reset rate becomes too high, this balance is reversed, and a minimum habitat size emerges at that tipping point.

The results are summarized in the diagram of Fig.~\ref{fig:criticalline}(b), where we present, for each value of $a$, the line that delimits the region in the plane of parameters $x_r$--$r$, for which resetting reduces $\ell_c$, 
depicted by a dashed line when $x_r\in\Omega_{\rm out}$ or by a solid line when $x_r\in\Omega_{\rm in}$. 
Moreover, for $a=0.1$,  
we distinguish the regions where the reduction occurs,
depending on whether $x_r$ lies inside (double-hashed) or outside (hashed) the patch. 
The line that delimits the two regions is obtained by setting $\ell_c=x_r$ in Eq.~(\ref{eq:criticalell-all2in}), or alternatively in Eq.~(\ref{eq:condition-all2out}), yielding 

\begin{equation} \label{eq:lc=rr-all}
     x_r=\frac{1}{\sqrt{1-r}}\arccos\sqrt{\frac{(a+r)}{
     (a+1)[ r+a(1-r) ] }}, 
\end{equation}
as far as $\ell_c<\ell_c^*$.
Separate diagrams for three individual different values of $a$ are displayed in 
Appendix~\ref{app:diagrams} (Fig.~\ref{fig:diagrams}).

%%%%
\subsection{Partial relocation }
\label{subsec:out2*}

In this case, only organisms localized in the external region $\Omega_{\rm out}$ are repositioned. In such a case $\varepsilon=0$, and the stationary form of Eqs.~(\ref{eq:u}) become
\begin{align}
 & u'' +  u + r N_{ \Omega_{\rm out}} \,\delta(x-x_r)=0,  & \quad & \mbox{if } x\in \Omega_{\rm in},\nonumber\\
 &  u'' - (a+ r)u+r N_{ \Omega_{\rm out}} \,\delta(x-x_r)=0,                              & \quad & \mbox{if } x\in \Omega_{\rm out}.
   \label{eq:out2*}   
\end{align} 
Following the same approach as in Sec~\ref{subsec:all2*}, if $x_r\in \Omega_{\rm in}$, that is, if organisms are relocated to the patch, then the critical size is given by
\begin{equation}
    \ell_c(r,x_r)=   \arccos
    \Biggl(
    \frac{  r\cos{x_r} +\sqrt{\Lambda'}}
    {1+ a+r }
    \Biggr),
    \label{eq:criticalell-out2in}
\end{equation}
where 
$\Lambda'=1+ a+r -  r^2 \cos^2{x_r}/(a+r)$.

If relocations are from outside the patch, i.e., $x_r\in \Omega_{\rm out}$, then the critical half-width is obtained by numerically solving the equation
\begin{equation} \label{eq:condition-out2out}
{\rm e}^{\gamma x_r}(\gamma^2-r)(\gamma    \cos\ell_c -\sin\ell_c) - {\rm e}^{\gamma\ell_c}\sin\ell_c = 0,
\end{equation}
where $\gamma = \sqrt{a+r}$.

%%%
 Figure~\ref{fig:finite-out}(a) shows the critical half-width 
$\ell_c$ as a function of $r$, for $x_r=0$ and different values of $a$. 
The picture is qualitatively similar to that observed in Fig.~\ref{fig:finite-all}(a). However, in this case the reset mechanism is less effective in reducing the critical size, as can be seen by the slower decay and lower ratio $\ell_c^*/\ell_c(\infty,x_r)$ for the same value of $a$. In the limit $a\to\infty$, $\ell_c$ becomes independent of $r$, in contrast to the case of total relocation.
Note that only individuals in the hostile region are ``rescued'' in the present case.

\begin{figure}[h!]
    \centering
\includegraphics[width=0.45\textwidth]{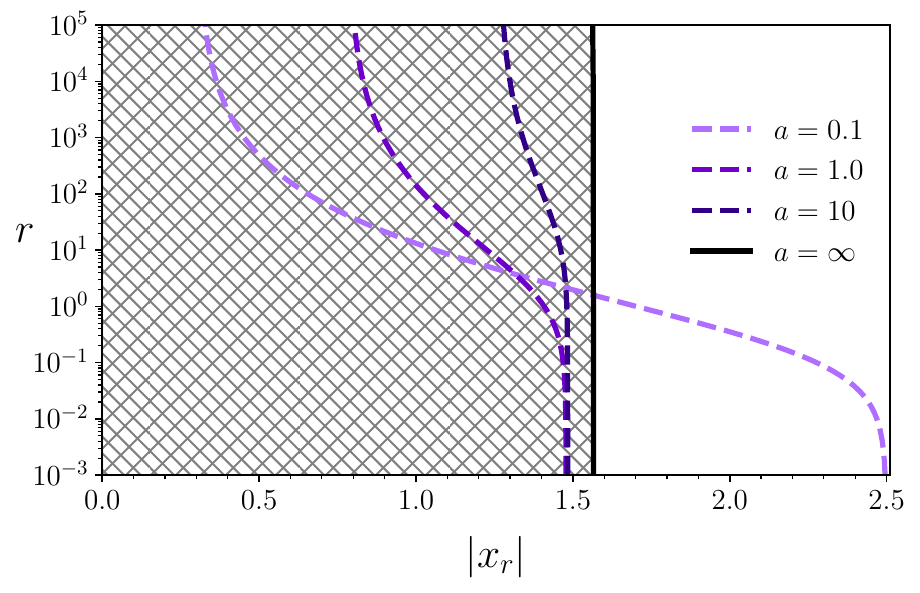}
    \caption{
    Diagram for partial relocation in the space of parameters $(x_r,r)$ showing the lines that delimit the region (to the left of the curves) where resetting reduces the critical patch size, given by the condition
$\ell_c(r,x_r)=\ell_c^*$, as in Fig.~\ref{fig:criticalline}(b).
In the case $a \to \infty$, the double-hashed region to the left of the line $x_r=\ell_c^*=\pi/2$ allows reduction of the critical size for any rate, which occurs for $x_r\in\Omega_{\rm in}$. 
For different values of $a$ the curve  $r^*$ vs.\ $|x_r|$ is displayed. }
    \label{fig:diagram2}
\end{figure}

When changing $x_r$, new nonmonotonic behaviors of $\ell_c$ vs.~$r$ emerge, as shown in Fig.~\ref{fig:finite-all}(b). 
Even when the reset position is outside the patch (dashed lines), resetting can be beneficial. 
Furthermore, for certain values of $x_r$ (outside the patch), there is an optimal rate for which $\ell_c$ is minimal (as illustrated in the inset for $x_r=1.3$), smaller than in the reset-free case. The appearance of this optimal value can be understood by similar arguments to those used in relation to the optimal values in Fig.~\ref{fig:finite-all}. 
The full portrait is schematized in Fig.~\ref{fig:diagram2}, 
where we plot, for each value of $a$ in the legend, 
the curve that defines the region where reset reduces the critical size (to the left of the curve).  
 
The line that delimits the two regions is obtained by setting $\ell_c=x_r$ in Eq.~(\ref{eq:criticalell-out2in}), or alternatively in Eq.~(\ref{eq:condition-out2out}), yielding 

\begin{equation} \label{eq:lc=rr-partial}
     x_r=\arccos\sqrt{\frac{(a+r)(1+a+r)}{ (a+r)(a+1)^2+r^2 }}.
\end{equation}

Also in this case, diagrams for three values of $a$ are displayed in 
Appendix~\ref{app:diagrams} (Fig.~\ref{fig:diagrams}), where we distinguish the region where resetting diminishes the critical size, for reset positions $x_r$ either outside (hashed) or inside (double hashed) the habitat.

\section{Conclusions and final remarks}
\label{sec:final}

We studied the effects of stochastic resetting in the critical patch size problem. 
Organisms can be subject to stochastic spatial repositioning into or to the outside of a viable patch. We first focused on the specific scenario of an extremely hostile environment beyond the boundaries of a good-quality region. 
We obtained a robust analytical approximation to describe the dynamics of the total population $N$ over time, providing insight into the long-term behavior of the population in such extreme environments.
Furthermore, we analytically derived an explicit expression for the critical patch size, $2\ell_c$, as a function of the parameters of the population dynamics and stochastic resetting.  

Next, we extended our study to address a more general scenario, where the environment outside the habitat is not entirely lethal but is characterized by a finite mortality rate. This scenario is particularly relevant for real-world ecosystems, where habitats often exist in a complex mosaic of varying environmental conditions. We elaborated the cases where organisms can be relocated from the outside or from anywhere, into the patch or outside it. In some cases, we arrive at a closed form expression for the critical habitat size, while in other ones we obtained a transcendental equation from which $\ell_c$ can be numerically extracted.  

In some cases, we observe that the critical size can decrease with resetting, the more the higher are the reset rates. In other cases, depending on $x_r$, there may also be a pessimal scenario in which there is a reset rate that makes the critical size reach its maximum value. These regimes were clearly delineated using our analytical results.
Similar results are obtained for the partially hostile environment, when $a$ is large. However, for small $a$, a minimal $\ell_c$ emerges at a finite optimal rate $r$ when organisms relocate to the outer region.

These results deepen our understanding of how environmental structure and stochastic sporadic events jointly influence population persistence in fragmented or degraded landscapes. One key consequence of our findings is the identification of non-monotonic dependencies between resetting rate and critical habitat size, revealing that movement strategies based on intermittent relocation can either facilitate survival or exacerbate extinction risks, depending on both the reset location and the environmental hostility. This challenges the classical intuition that more frequent returns to a safe location always enhance survival, highlighting instead the subtle interplay between movement dynamics and spatial context. As future perspectives, a whole research line becomes possible, such as considering scenarios with heterogeneous environments\cite{DOSSANTOS2020,PhysRevE.110.054111}, density-dependent processes~\cite{COLOMBO2018,PIVA2025}, non-instantaneous resetting~\cite{bodrova2020resetting,mercado2020intermittent,gupta2021confining} or considering a distribution of reset positions.

\section*{Acknowledgments}
We all acknowledge partial financial support by the 
Coordena\c c\~ao de Aperfei\c coamento de Pessoal de N\'{\i}vel Superior
 - Brazil (CAPES) - Finance Code 001. C.A. also acknowledges the partial financial support (311435/ 2020-3) of 
 Conselho Nacional de Desenvolvimento Cient\'{\i}fico e Tecnol\'ogico (CNPq), Brazil, 
and (CNE E-26/204.130/2024) Funda\c c\~ao de Amparo \`a Pesquisa do Estado do Rio de Janeiro (FAPERJ), Brazil. P.d.C.\ was supported by Scholarships No.\ 2021/10139-2 and No.\ 2022/13872-5 and ICTP-SAIFR Grant No.\ 2021/14335-0, all granted by São Paulo Research Foundation (FAPESP), Brazil.

\bibliography{ref}

%\onecolumngrid

\appendix
\section{Solving Eq.~(\ref{eq:N_diff})}
\label{app:timesolution}

We proceed to solve Eq.~(\ref{eq:N_diff}) by Laplace
transforming in time, which yields
\begin{equation}
    \partial^2_{x_0} N(x_0,s)+ (1-r-s)N(x_0,s) + r N(x_r,s) + N_0=0, 
    \label{eq:N_dif_laplace}
\end{equation}
where $N_0$ is the initial population. The general solution of Eq.~(\ref{eq:N_dif_laplace}) is 
\begin{eqnarray}
    N(x_0,s) &=& c_1 \cos( x_0 \sqrt{1-r-s} ) + c_2 \sin(x_0 \sqrt{1-r-s}  )\nonumber  \\&-& \frac{r N(x_r,s)+N_0}{1-r-s}, 
    \label{eq:solution_N}
\end{eqnarray}
where functions are defined for complex argument.

Since  $N$ vanishes for the initial conditions $x_0=\pm \ell$, then,   $c_2=0$ and 
\begin{equation}
     c_1=\frac{r N(x_r,s)+N_0}{1-r-s} \sec( \ell \sqrt{1-r-s}  ).
 \end{equation}
Furthermore, Eq.~(\ref{eq:solution_N}) must be solved self-consistently, by setting $x_0 = x_r$, which yields  
\begin{equation}
    N( x_0,s) = N_0 \frac{\cos( x_0 \xi )-\cos(  \ell  \xi)}{(1-s)\cos(  \ell  \xi)-r\,\cos( x_r \xi)},
    \label{eq:N_in_laplace}
\end{equation}
with $\xi = \sqrt{1-r-s}$.

Finally, Laplace transform inversion can be obtained by the Mellin integral
\begin{equation}
      \frac{N(x_0,t)}{N_0} = \int_{\gamma -i \infty}^{\gamma + i \infty} \frac{ds \,e^{st}}{2\pi i} \frac{\cos( x_0 \xi )-\cos(  \ell  \xi)}{(1-s)\cos(  \ell  \xi)-r\,\cos( x_r \xi)},
      \label{eq:melin_int}
\end{equation}
where $\gamma$ is a suitable real constant~\cite{butkov1973mathematical}. The integrand in Eq.~(\ref{eq:melin_int}) exhibits a singularity structure characterized by poles $s^*$ determined as the solutions of
\begin{equation}
   ( 1-s^*)\cos\left[\ell\sqrt{1-r-s^*}\right]-r\cos( x_r \sqrt{1-r-s^*}) =0, 
   \label{eq:poles}
\end{equation}
which are real valued and can be obtained numerically.  
In the absence of reset ($r=0$),  Eq.~(\ref{eq:poles}) yields $s^*= 1-[(2n+1)\pi/(2\ell)]^2$, with $n=0,1,\ldots$, and Eq.~(\ref{eq:melin_int}) recovers the known result~\cite{skellam1991random,ludwig1979spatial}, namely   
$  
N(t)=  \frac{4N_0}{\pi}\sum_{n=0}^\infty \frac{(-1)^n }{2n + 1} \exp\left[ \left(1 - \left[(2n+1)\pi/(2\ell)\right]^2\right)t \right ] 
$.

For large $t$, the dominant contribution to inversion will be determined by the largest pole $s_M={\rm max}\{s^*\}$. By applying Cauchy's residue theorem for this pole, we obtain
\begin{equation}
       N(t) \approx {\cal N} \,e^{s_M t},
      \label{eq:exp}
\end{equation}
where ${\cal N}\equiv {\cal N}(x_0,x_r,r, \ell)$ is given by
\begin{equation}
{\cal N} =  \frac{4 N_0\,\xi_r\left[ \cos( \ell  \, \xi_r) - \cos(x_0 \, \xi_r)  \right]}{ 4 \,\xi_r \cos( \ell \, \xi_r) - \ell\,( 1-s_M )\sin( \ell \, \xi_r) + 2r\,x_r \sin{(x_r\xi_r)} },
  \label{eq:pre_factor0}
\end{equation}
where $\xi_r = \sqrt{1-r-s_M}$ and $N_0$ is the initial total population.

In this way, when $s^*=0$, Eq.~(\ref{eq:poles}) straightforwardly provides the critical value of $\ell$, namely,  Eq.~(\ref{eq:criticalell-xr}).

\section{Solving the stationary form of Eq.~(\ref{eq:N_diff})}
\label{app:steadysolution-infty}

An alternative approach to deriving $\ell_c$ 
consists in solving the stationary form of Eq.~(\ref{eq:N_diff}), namely, 
\begin{equation}
\partial^2_{x_0} N(x_0) + (1-r)N(x_0)  + r N(x_r)=0,  
    \label{eq:N_diff-ss}
\end{equation}
with the boundary conditions  $N(x_0=\pm \ell_c )=0$.

For $r\neq 1$, we obtain
\begin{equation} \label{eq:Nx0}
    N(x_0)= \frac{r \,N(x_r)}{1-r}  \left( \frac{\cos(x_0\sqrt{1-r})}{\cos( \ell_c \sqrt{1-r})} -1 \right). 
\end{equation}
By setting $x_0=x_r$, for consistency, we directly obtain, in an alternative way, Eq.~(\ref{eq:criticalell-xr}). 
 
For $r=1$, the solution of Eq.~(\ref{eq:N_diff-ss}) that satisfies the boundary conditions is  
\begin{equation}
    N(x_0)= N(x_r)[  \ell_c ^2- x_0^2]/2, 
\end{equation}
which, from the consistency relation at $x_0=x_r$ gives
 \begin{equation}
     \ell_c = \sqrt{x_r^2 + 2}, 
 \end{equation}
which can also be obtained by taking the limit $r\to 1$ in Eq.~(\ref{eq:Nx0}).

\section{Stationary solution for a partially hostile environment}
\label{app:steadysolution-all}

In all cases, to address the stationary problem,  we begin by setting Eqs.~(\ref{eq:u}) equal to zero. Owing to their linear structure, the solutions in each region take the form of real exponential functions in the outer domains, and a combination of sine and cosine functions (either with real or imaginary arguments, depending on the parameters) within the patch.

The solutions must verify continuity at $x=\pm \ell_c$ and at $x=x_r$. Additionally, the flux must be continuous at $x=\pm \ell_c$, which imposes $u'_{\rm out}=u'_{\rm in}$. 
At $x=x_r$, the presence of a Dirac delta introduces a discontinuity in the derivative, leading to a jump condition 
  $u'(x_r^+)-u'(x_r^-)=-r\,N_{\Omega}$.  
These six conditions lead to a homogeneous system of equations for the coefficients. To obtain nontrivial solutions, the determinant of the resulting coefficient matrix must vanish. This condition determines the value of $\ell_c$.
   
The procedure is illustrated for the case of Eq.~(\ref{eq:out2*}). 
When $x_r\in\Omega_{\rm in}$, the stationary solution is
\begin{equation}
  u(x) = 
\begin{cases} 
     c_1 e^{\gamma x }, & x \in (-\infty,-\ell_c ], \\
     c_2 \sin( x) + c_3\cos( x) & x\in[-\ell_c,x_r],\\
     c_4 \sin(  x) + c_5\cos( x) & x\in[x_r,\ell_c],\\
     c_6 e^{-\gamma x} & x\in[\ell_c,\infty ),  
\end{cases}
\label{eq:u-out2in}
\end{equation}
with $\gamma=\sqrt{a+r}$. The continuity and boundary conditions explicitly are
\begin{eqnarray*} \label{eq:cond}
    u_1(-\ell_c) &=& u_2(-\ell_c)\\
    u'_1(-\ell_c) &=& u'_2(-\ell_c)\\
    u_2(x_r) &=& u_3(x_r) \\
      u'_3(x_r) -u'_2(x_r) &=& -r N_{\rm out} \\
     u_3(\ell_c) &=& u_4(\ell_c)\\ 
       u'_3(\ell_c) &=& u'_4(\ell_c),
\end{eqnarray*}
where $N_{\rm out}=\int_{-\infty}^{\ell_c} u_1(x) dx+
\int_{\ell_c}^\infty  u_4(x) dx.$ 

Using Eq.~(\ref{eq:u-out2in}), these conditions can be written in matrix form as $\mathbf{M}(c_1,c_2,\ldots,c_6)^T =(0,0,\ldots,0)^T$, 
where
\begin{equation}
\footnotesize
\mathbf{M}=\begin{pmatrix}
{\rm e}^{-\gamma \ell_c}  &\sin\ell_c   &-\cos\ell_c &0 &0 &0\\
 \gamma{\rm e}^{-\gamma \ell_c} & -\cos\ell_c &-\sin\ell_c &0 &0 &0\\
0&  \sin x_r &\cos x_r &-\sin x_r &-\cos x_r &0\\
\frac{r}{\gamma}{\rm e}^{-\gamma \ell_c} 
& - \cos  x_r  &  \sin  x_r   & \cos  x_r  &- \sin x_r 
&\frac{r}{\gamma}{\rm e}^{-\gamma \ell_c}\\
0 &0  &0 &\sin \ell_c &\cos \ell_c &-{\rm e}^{-\gamma \ell_c}\\
0 &0  &0 & \cos \ell_c  &- \sin \ell_c &\gamma {\rm e}^{-\gamma \ell_c}
\end{pmatrix}  
\label{eq: matriz_sis}
\end{equation}
Since the system is homogeneous, a nontrivial solution, requires $\det\mathbf{M}=0$, which leads to the equation
\begin{equation}
   \gamma^2\cos \ell_c -r \cos x_r - \gamma \sin \ell_c  =0,\
\end{equation}
from which $\cos \ell_c$ can be extracted, yielding Eq.~(\ref{eq:criticalell-out2in}).
 
For the case $x_r\in \Omega_{\rm out}$, the solution of Eq.~(\ref{eq:out2*}) has the form
\begin{equation}
  u(x) = 
\begin{cases} 
     c_1 e^{\gamma x }, & x \in (-\infty,-\ell_c ], \\
     c_2 \sin( x) + c_3\cos( x) & x\in[-\ell_c,\ell_c],\\
     c_4 e^{\gamma x } + c_5e^{-\gamma x } & x\in[\ell_c,x_r],\\
     c_6 e^{-\gamma x} & x\in[x_r,\infty ).   
\end{cases}
\label{eq:u-out2out}
\end{equation}
Applying the corresponding continuity and boundary conditions to produce the matrix $\mathbf{M}$ (not shown), 
the condition of null determinant leads to the  transcendental Eq.~(\ref{eq:condition-out2out}), which was solved numerically. 

We proceed analogously in the cases where repositions can occur from any place. The main difference is that the argument for the $\sin$ and $\cos$ functions
is $\alpha x$ (which for $r>1$ leads to hyperbolic functions), and that the jump in the derivative due to the Dirac delta is proportional to the total $N$, instead of $N_{ \Omega_{\rm out}}$.
Then, for $x_r \in \Omega_{\rm in}$, the matrix singularity condition leads to
\begin{equation}
    - \gamma^2 \cos(\alpha \ell_c) + 
 r (1-r + \gamma^2) \cos(\alpha x_r) + \alpha \gamma \sin(\alpha \ell_c)=0,
\end{equation}
which can be solved for $\cos(\alpha \ell_c)$ to obtain the closed form Eq.~(\ref{eq:criticalell-all2in}),
while for $x_r \in \Omega_{\rm out}$, the matrix singularity condition yields the transcendental Eq.~(\ref{eq:condition-all2out}).

\newpage
\widetext
\section{Phase diagrams}
\label{app:diagrams}
\begin{figure*}[h!]
    \centering
\includegraphics[width=0.32\textwidth]{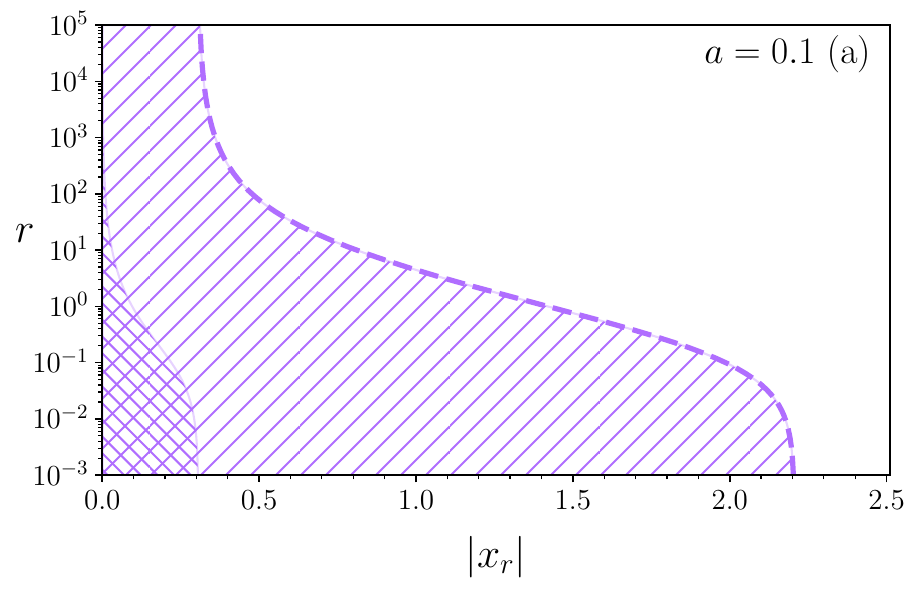}
\includegraphics[width=0.32\textwidth]{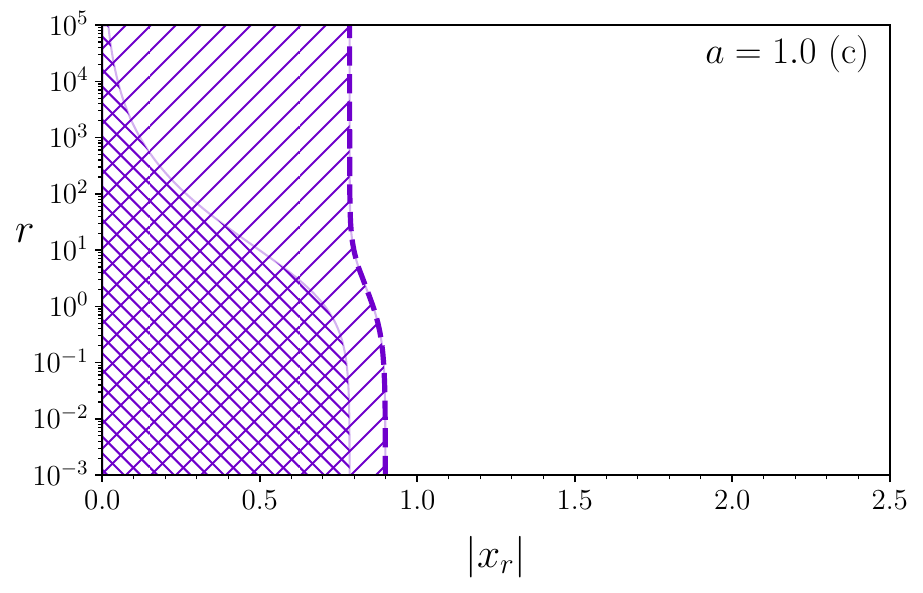}
\includegraphics[width=0.32\textwidth]{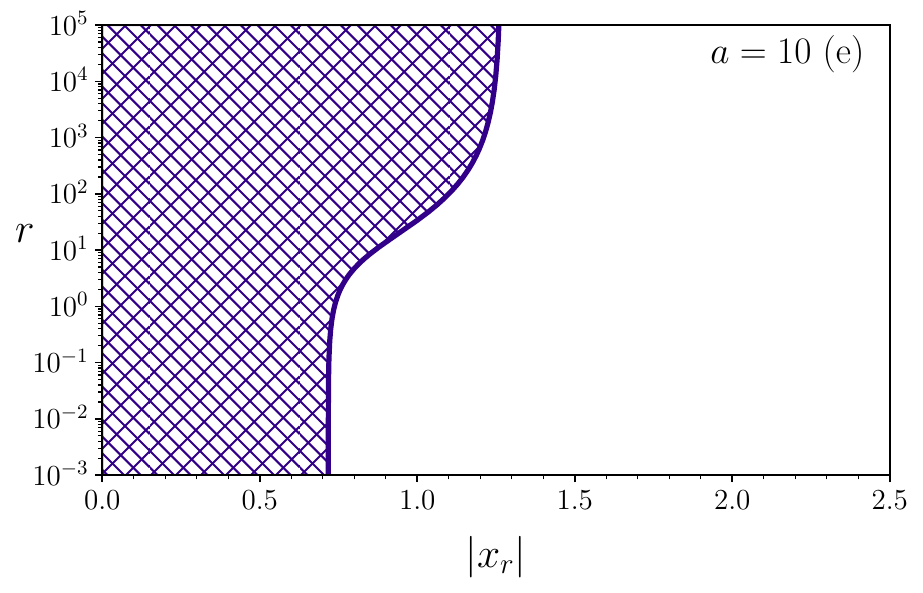}
\includegraphics[width=0.32\textwidth]{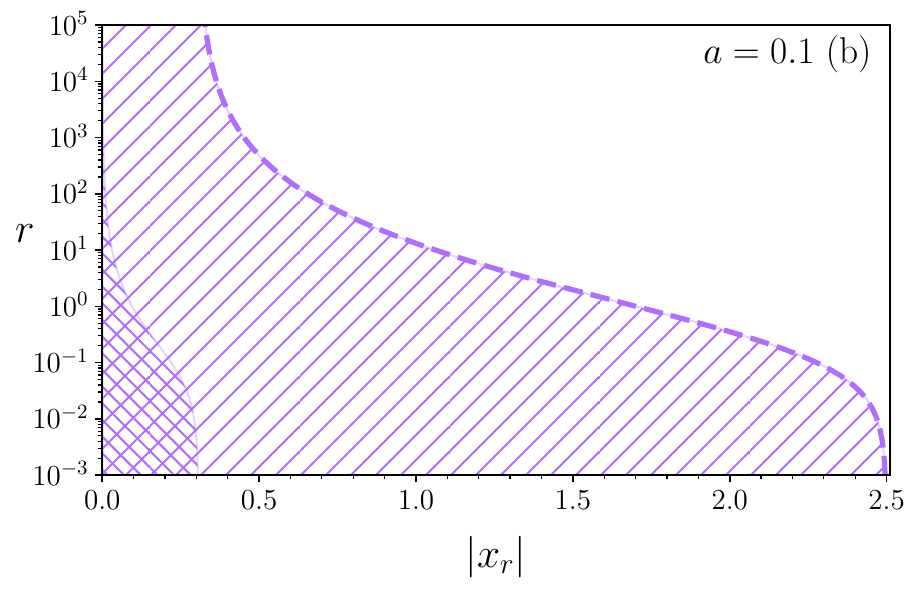}
\includegraphics[width=0.32\textwidth]{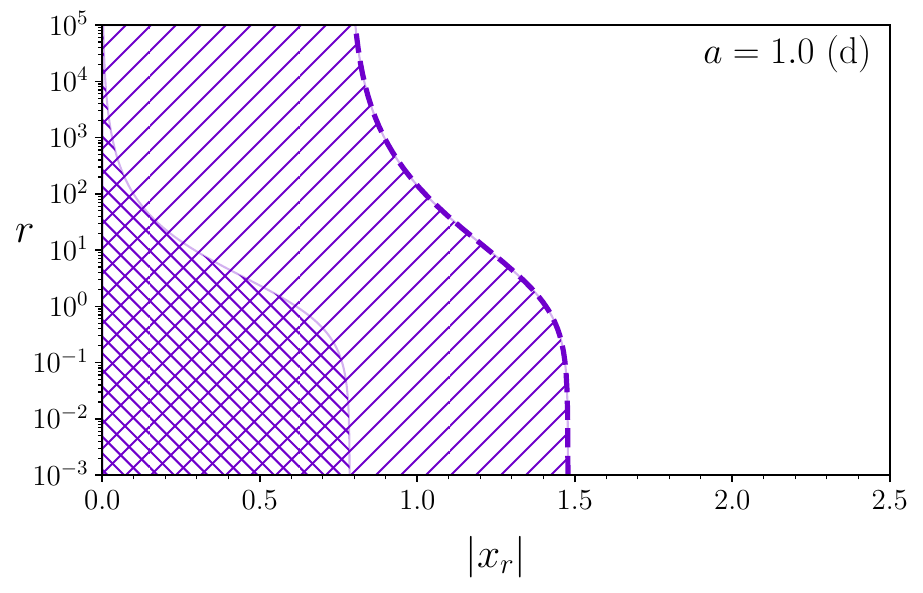}
\includegraphics[width=0.32\textwidth]{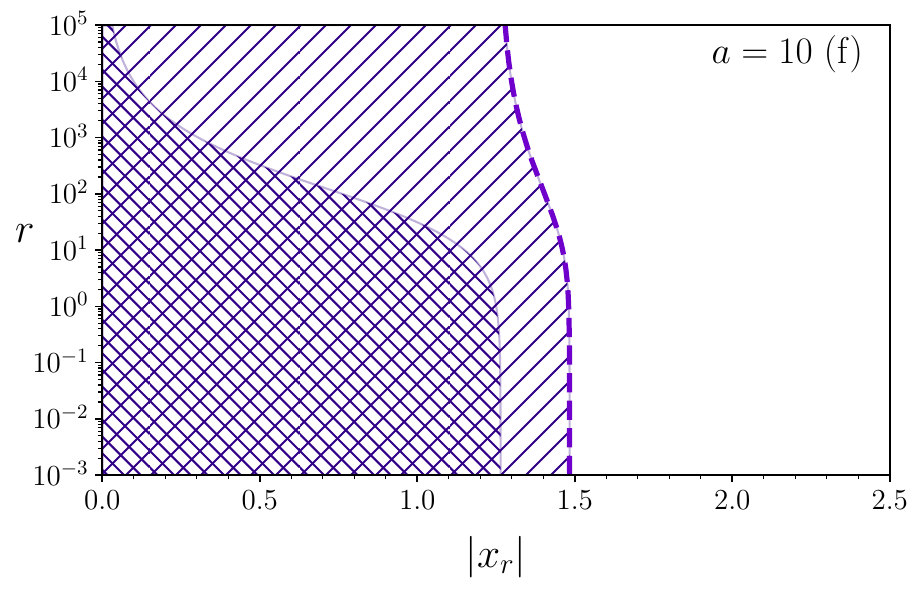}
    \caption{Diagrams in the plane $x_r$--$r$, 
    for total relocation, with $\Omega=\Omega_{\rm in} \cup \Omega_{\rm out}$ (upper row), and partial relocation, with
    $\Omega= \Omega_{\rm out}$ (lower row).
    Each of the three distinct values of $a$ corresponds to a column, with colors corresponding to those of Figs.~\ref{fig:finite-all}(a) and \ref{fig:finite-out}(a).
The shadowed region is where resetting reduces the critical patch size, delimited by the condition $\ell_c(r,x_r)=\ell_c^*$, depicted by a dashed line when $x_r\in\Omega_{\rm out}$ or a solid line when $x_r\in\Omega_{\rm in}$. 
The double-hashed (or hashed) region corresponds to critical size reduction by resetting when $x_r$ belongs (or not) to the patch, which are delimited by the condition $\ell_c=x_r$, given that $\ell_c<\ell_c^*$ (thin solid line).   
}
    \label{fig:diagrams}
\end{figure*}

\end{document}